\documentclass{iau} 
\usepackage{graphicx}
\usepackage{titlesec}
\titlespacing\section{0pt}{15pt plus 4pt minus 2pt}{1pt plus 2pt minus 2pt}

\title[IAU 337.~~Pulsar science with CHIME] 
{Pulsar science with the CHIME telescope}

\author[Cherry Ng et al.]   
{Cherry Ng$^{1,2}$
 on behalf of the CHIME Pulsar collaboration}

\affiliation{$^1$Dunlap Institute, University of Toronto, \\ 
50 St. George St., Toronto, ON~M5S~3H4, Canada \\ email: {\tt cherry.ng@dunlap.utoronto.ca} \\[\affilskip]
$^2$Dept. of Physics and Astronomy, University of British Columbia, \\
6224 Agricultural Road, Vancouver, BC~V6T~1Z1, Canada\\}

\pubyear{2017}
\volume{337}  
\jname{Pulsar Astrophysics -­ The Next 50 Years}
\editors{P. Weltevrede, B.B.P. Perera, L. Levin Preston \& S. Sanidas, eds.}

\begin{document}

\maketitle

\begin{abstract}
The CHIME telescope (the Canadian Hydrogen Intensity Mapping Experiment) recently 
built in Penticton, Canada, is currently being commissioned. Originally 
designed as a cosmology experiment, it was soon recognized that CHIME has the potential 
to simultaneously serve as an incredibly useful radio telescope for pulsar science. CHIME operates 
across a wide bandwidth of 400$-$800\,MHz and will have a collecting area and 
sensitivity comparable to that of the 100-m class radio telescopes. 
CHIME has a huge field of view of $\sim$250 square degrees.
It will be 
capable of observing 10 pulsars simultaneously, 24-hours per day, every day, 
while still accomplishing its missions to study Baryon Acoustic Oscillations and Fast Radio Bursts. It will 
carry out daily monitoring of roughly 
half of all pulsars in the northern hemisphere, 
including all NANOGrav pulsars employed in the Pulsar Timing Array project.
It will cycle through all pulsars in the northern hemisphere with a range of cadence of no more than 10 days. 

\end{abstract}

\firstsection 
\section{Telescope overview}
The Canadian Hydrogen Intensity Mapping Experiment (CHIME\footnote{www.chime-experiment.ca}) 
is a radio telescope recently constructed
at the Dominion Radio Astrophysical Observatory (DRAO) 
in Penticton, BC, Canada, and which is currently being commissioned.
CHIME is composed of four cylindrical reflecting 
surfaces, each 100-m in length North-South (N-S) and 20-m wide East-West (E-W). 
This geometry provides an extremely wide effective 
field-of-view (FOV) of $\sim$120$^{\circ}$ in N-S and 
1.3$-$2.5$^{\circ}$ (frequency dependent) in E-W, that is, a primary beam size of $\sim$250 square degrees. 
With no moving parts in the structure of CHIME, it operates as a transit telescope, 
surveying the entire overhead sky each day as the earth rotates.
On each of the four focal lines is a linear array of 256 dual polarization antennas.
These antennas are made using printed circuit boards and have a clover-leaf shape that  
optimizes for the broad bandwidth of CHIME from 400-800\,MHz 
(\cite[Deng et al. 2014]{Deng2014}).
These antennas are arranged in a regular grid with a mean spacing of 0.3048\,m 
in the N-S direction and 22\,m in E-W. 
We thus have a total of 2048 analog inputs and these are each amplified and brought 
through 50\,m-long co-axial cables to 
a digital F-X correlator.

\begin{figure}[htbp]
  \centering
  \includegraphics[width=5in]{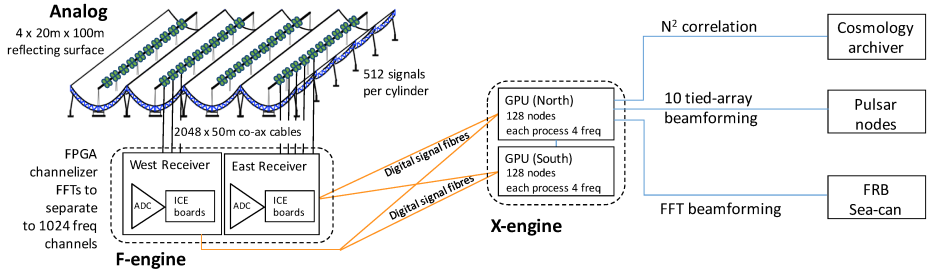}
  \caption{A system diagram of CHIME, showing the signal path from the 
analog system to the F-engine, X-engine, and the three independent data 
backends.}
\vspace{-12px}
  \label{fig:system}
\end{figure}

In the correlator F-engine, 128 custom-made signal processing boards 
based on field programmable gate arrays (FPGAs; \cite[Bandura et al. 2016a]{Bandura2016a},
\cite[2016b]{Bandura2016b}) digitize the 
analog radio signals collected and channelize the full bandwidth into 1024 
frequency bins via a 4-tap polyphase filter bank (PFB), at a 2.56$\,\mu$s cadence 
(see Table \ref{tab:spec}). 
In the X-engine, spatial correlation is performed in a GPU 
cluster that consists of 256 processing nodes each with 4 AMD Fiji GPUs,
building on the Pathfinder system described in
\cite[Recnik et al. (2015)]{Recnik2015}, 
\cite[Denman et al. (2015)]{Denman2015}, and 
\cite[Klages et al. (2015)]{Klages2015}. This GPU cluster 
also forms 10 dual-polarization tied-array beams, 
allowing us to track 10 
pulsars at different locations simultaneously throughout the primary beam.
A 10-node GPU-based pulsar backend will 
create folded pulsar archives 
with coherent dedispersion using 
the \textsc{dspsr} software 
(\cite[van Straten et al. 2011]{DSPSR}). The output data will have 1024 frequency channels and 
1024 phase bins with 4 stokes at a bit depth of 16. We will have a read out cadence of 10\,s which 
means a total output rate of 67\,Mbps.
See Figure~\ref{fig:system} for a system diagram of CHIME.

\begin{table}
\begin{center}
  \caption{The specifications of the CHIME/Pulsar project.}
{\scriptsize
  \begin{tabular}{ll|ll}\hline
\multicolumn{2}{c|}{CHIME general parameters} & \multicolumn{2}{c}{Pulsar-backend specific parameters} \\
\hline
Field of view & 120$^{\circ}$ (N-S) ; 1.3-2.5$^{\circ}$ (E-W) = 250\,sq degree & 
Number of phase bins & 1024 \\
Beam size & 0.26$^{\circ}$ at 800\,MHz; 0.52$^{\circ}$ at 400\,MHz & 
Frequency resolution & 390\,kHz \\
System temperature & $\sim$50\,K  & 
Number of spectral channels & 1024 \\
Bandwidth & 400-800\,MHz &
Output data bit depth & 16 bits\\
Telescope latitude & 49$^{\circ}$19.2\,m & Number of polarizations & 2 \\
Telescope longitude  & $-$117$^{\circ}$37.2' & Pulsar data output rate & 67\,Mbps\\

\hline \label{tab:spec}
 \end{tabular}
}
 \end{center}
\vspace{-0.8\skip\footins}
\end{table}

\section{Telescope sensitivity}
\begin{figure}[htbp]
  \centering
  \includegraphics[width=4.6in]{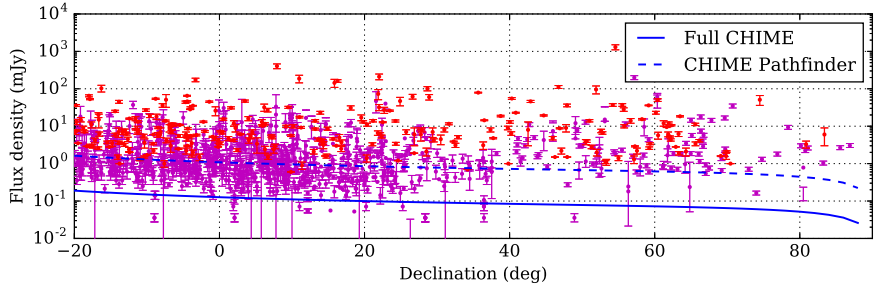}
  \caption{Minimum detectability of CHIME (solid line) and the Pathfinder (dashed line) as determined 
from the radiometer equation. Pulsars above $\delta$=$-$20$^{\circ}$ with 
published \textsc{psrcat} flux density at 600\,MHz are plotted as red dots and the 
error bars represent 1\,$\sigma$ uncertainties. For pulsars 
where only the flux density at another observing frequency is available,
we scaled them to 600\,MHz using their respective spectral indices. 
These pulsars are plotted as magenta points.}
  \label{fig:flux}
\end{figure}

CHIME will be able to observe pulsars down to 
a declination of about $-$20$^{\circ}$ (see Table~\ref{tab:spec}). 
Because CHIME is a transit telescope, the maximum dwell time on each pulsar 
is a function of the source declination ($\delta$). At the equator, this is approximately 
10-15\,minutes, while within the circumpolar region a source can be tracked 
for hours. 
Figure~\ref{fig:flux} shows the minimum detectable flux density of CHIME 
and the CHIME Pathfinder (\cite[Bandura et al. 2014]{Bandura2014}), with the above factors taken 
into account. 
These limits are compared to 
the flux density of known pulsars above $\delta$=$-$20$^{\circ}$ 
published in the 
\textit{ATNF Pulsar Catalogue}\footnote{http://www.atnf.csiro.au/people/pulsar/psrcat/} 
(\cite[Manchester et al. 2005]{PSRCAT}). It can be seen 
that CHIME should be able to detect the majority  
of the known pulsars in its visible sky.

\section{Science case}
Apart from the original science goal of mapping redshifted 21-cm hydrogen emission 
to study the Baryon Acoustic Oscillation signal at redshifts 0.8-2.5 
(\cite[Bandura et al. 2014]{Bandura2014}), 
as well as the search of Fast Radio Bursts (FRBs; see, e.g., \cite[Ng et al. 2017]{Ng2017}), 
a third CHIME backend is being deployed for high cadence pulsar timing, the focus of this proceeding.

\subsection{Pulsar timing for Gravitational wave detection}
CHIME plans to observe all visible NANOGrav\footnote{http://nanograv.org/} pulsars 
daily to aid the effort of 
Pulsar Timing Array (PTA) experiments 
to detect Gravitational waves. 
Temporal dispersion measure (DM) and scattering variations will be monitored daily for each NANOGrav 
pulsar. 
The upper part of the CHIME band overlaps with the NANOGrav 820-MHz observing band at the 
Green Bank Telescope (GBT). 
CHIME's wide bandwidth will provide a large `lever arm' for
measuring the DM as well as scattering time.  In principle, the 
CHIME-measured, daily DMs could be used to reduce interstellar
medium-related noise in NANOGrav data.
Simulations suggest that this could mean an improvement in timing precision
by a factor of at least 2 in at least half of the NANOGrav pulsars.
However, \cite[Cordes et al. (2016)]{Cordes2016} argue that DM variations 
might be effectively a frequency-dependent phenomenon, and hence the 
DM observed by CHIME at low frequencies would be 
distinct from that relevant at the higher PTA observing frequencies.
CHIME will conduct regular simultaneous 
observations in conjunction with the GBT and the Arecibo Telescope at 
higher radio frequencies, which will help test models of the interstellar medium.
CHIME will be of great use not only for NANOGrav but for all international PTA projects.

\subsection{Pulsars with time domain variability}
Not all pulsars have the same level of stability as those being employed in PTA experiments. 
Some pulsars exhibit time domain variability. 
For example, glitches in pulsars are discrete changes of the pulsar rotation rate thought to be a 
probe of the neutron-star interior. In one study,
\cite[Espinoza et al. (2011)]{Espinoza2011} showed that 
thus far, 482 glitches have been observed from 
168 pulsars\footnote{See the online glitch table at http://www.jb.man.ac.uk/pulsar/glitches.html}, although it is 
believed that the fraction of glitching pulsars as well as the frequency of glitch occurrence should be much higher.
As another example, \cite[McLaughlin et al. (2006)]{McLaughlin2006} discovered 
an entire class of pulsars with intermittent emission, 
now known as Rotating Radio Transients (RRATs).
Telescope time on sky 
is the greatest factor limiting a thorough follow up and complete census of these types of variable pulsars.
These sources will definitely benefit from a high observing cadence with CHIME.

CHIME also plans to perform daily monitoring of exotic objects such as 
high magnetic field (B$>$10$^{13}$\,G) pulsars which might one day show magnetar-like outbursts 
(e.g. \cite[Archibald et al. 2016]{Archibald2016}).
Monitoring will also be conducted on X-ray magnetars that are currently radio quiet 
and any radio loud magnetar in the CHIME sky. 
Transitional MSPs (tMSPs; see, e.g., 
\cite[Archibald et al. 2009]{Archibald2009}) swing between a rotation-powered state (radio emission) 
and an accretion-powered state (X-ray emission). 
CHIME will also be monitoring 
them to try and capture further moments of transition.

\section{Observing strategy}
\begin{figure}[htbp]
\centerline{\includegraphics[width=4.6in]{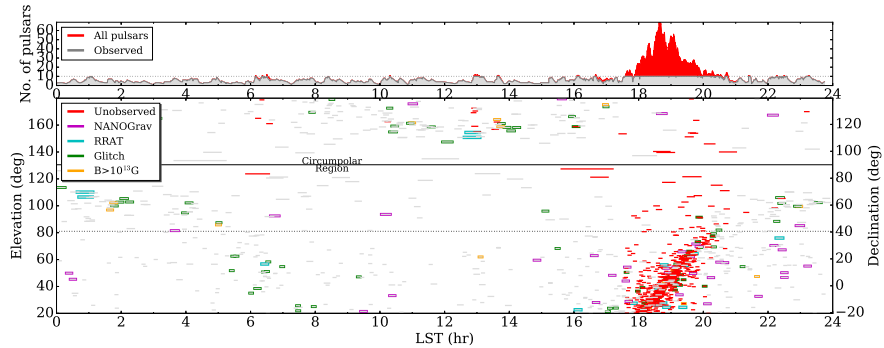}}
  \caption{The lower panel shows all pulsars over the CHIME sky and 
the length of each bar represents the drift time. Over the course of one 
day, CHIME will have observed all the gray bars, i.e., some 700 pulsars. 
Only along the Galactic plane will we have unobserved 
pulsars (red bars), but these sources 
will be prioritized and observed within the next 10 days.
The top panel shows the number of pulsars 
within the primary beam of CHIME at any given time.} 
\vspace{-15px}
  \label{fig:sky2}
\end{figure}

CHIME will have the ability to observe up to 10 pulsars 
simultaneously and with full polarization sensitivity, and is planned 
to operate 24/7 as pulsars transit overhead. 
This is revolutionary and according to simulations (Figure~\ref{fig:sky2}), 
will allow for daily observation for 
roughly half of all North-visible pulsars, including all NANOGrav pulsars.
CHIME will cycle through all pulsars in the northern hemisphere with a range of cadence no 
more than 10 days.
CHIME is in a unique position to provide us with extremely high cadence 
pulsar observations and will surely lead to exciting new insights in the field.

\section{Acknowledgements}
We are very grateful for the warm reception and skillful help we have received 
from the staff of DRAO, which is operated by the 
National Research Council of Canada. We acknowledge support from the Canada 
Foundation for Innovation, the Natural Sciences and Engineering Research Council of 
Canada, the B.C. Knowledge Development Fund, le Cofinancement gouvernement du Quebec-FCI, 
the Ontario Research Fund, the CIfAR Cosmology and Gravity program, the Canada Research 
Chairs program, and the National Research Council of Canada. We thank Xilinx University 
Programs for their generous support of the CHIME project, and AMD for donation of test 
units.

\end{document}